# High-quality multi-terminal suspended graphene devices

*Dong-Keun Ki and Alberto F. Morpurgo**

Départment de Physique de la Matiére Condensée (DPMC) and Group of Applied Physics (GAP), University of Geneva, 24 quai Ernest-Ansermet, CH-1211 Geneva 4, Switzerland.

We introduce a new scheme to realize suspended, multi-terminal graphene structures that can be current annealed successfully to obtain uniform, very high quality devices. A key aspect is that the bulky metallic contacts are not connected directly to the part of graphene probed by transport measurements, but only through etched constriction, which prevents the contacts from acting invasively. The device high quality and uniformity is demonstrated by a reproducibly narrow ($\delta n \sim 10^9$ cm$^{-2}$) resistance peak around charge neutrality, by carrier mobility values exceeding $10^6$ cm$^2$/V/s, by the observation of integer quantum Hall plateaus starting at 30 mT and of symmetry broken states at about 200 mT, and by the occurrence of a negative multi-terminal resistance directly proving the occurrence of ballistic transport. As these multi-terminal devices enable measurements that cannot be done in a simpler two-terminal configuration, we anticipate that their use in future studies of graphene-based systems will be particularly relevant.

KEYWORDS: Multi-terminal suspended graphene, Ballistic transport, Negative resistance, Graphene bilayer, Quantum Hall effect.

The investigation of the intrinsic electronic properties of graphene relies heavily on the study of transport through nano-electronic devices in which extrinsic disorder is minimized.[1-3] When produced through mechanical exfoliation of graphite, graphene and its multilayers usually exhibit an extremely high level of structural perfection and chemical purity, sufficient for the realization of very high-quality devices. However, eliminating the influence of external perturbations is very challenging because graphene-based materials have atomic scale thickness, so that nearby materials –the substrate or adsorbed molecules– can easily affect the carrier mobility and induce pronounced inhomogeneity in carrier density. Two main routes have been pursued to minimize these sources of extrinsic disorder. The first consists in realizing suspended devices,[1, 2] in which a graphene layer is hanging between source and drain electrodes without being in a direct contact with a substrate material. The second relies on the use of hexagonal boron-nitride (hBN) substrates,[3] which has been shown to enable the realization of high-quality graphene structures.

Key experiments have been performed by using either one of these techniques, with each having its own advantages and limitations. The realization of the graphene devices on hBN, for instance, allows the use of large area flakes, which enables the fabrication of more complex device structures.[4-8] It also appears that graphene on hBN can be cleaned from unwanted adsorbates rather reproducibly, once the device fabrication technique is under control.[9] However, it has been shown in a series of beautiful experiments[10-13] that, when placed onto hBN, charge carriers in graphene are influenced by the electrostatic potentials generated by the B and N atoms. As a result, owing to the lattice mismatch between graphene and hBN, the substrate causes a periodic modulation of the potential experienced by electrons in graphene, effectively generating an electronic superlattice.[10-13] The phenomena observed in the experiments imply that graphene on hBN possesses unique properties, different from those of intrinsic graphene. That is why it remains particularly important to investigate suspended graphene, which is unaffected by the influence of external electrostatic potentials.[1, 2, 14-23] Unfortunately, suspended graphene allows only rather small devices to be realized (larger devices are mechanically unstable), and even more stringent limitations on the experiments have been posed so far by the inability to controllably realize high-quality multi-terminal devices.[22, 24-27] To address this last issue, here we discuss a strategy to fabricate multi-terminal suspended devices with an electronic mean-free path exceeding the device size, and demonstrate the high quality of these devices through different transport measurements. In particular, we report the observation of a negative multi-terminal resistance due to the ballistic transport on suspended bilayer graphene, and very high quality quantum Hall data that include the independent determination of the transverse and the longitudinal resistance, as well as the observation of a pronounced quantum Hall plateau at magnetic field $B$ as low as 30 mT.

Before discussing the more technical aspects of our work, it is important to explain what difficulties have prevented so far the realization of high-quality multi-terminal suspended devices. It is not the fabrication of a multi-terminal geometry that poses problems, but the process to "clean" the suspended graphene layer from adsorbates.[22, 24] In two-terminal structures, cleaning is achieved by passing a very large current through the device.[1, 2] It is believed that the large temperature gradient generated during this "current annealing" step (the temperature reaches up to about 600 Celsius at the center of the device, and remains nearly unchanged close to the metal contacts)[28] induces diffusion of adsorbates from the central part of the layer towards the contacts, thereby "cleaning" the body of the suspended device. Whether the current annealing procedure has been successful can be easily determined by characterizing the device electrical properties, since properly cleaned graphene exhibits a very narrow resistance peak close to charge neutrality, high carrier mobility, etc.[1, 2, 14-23]

In multi-terminal geometries, however, it was found that passing a large current through graphene does not result in the systematic out-diffusion of adsorbates, and current annealing does not normally result in uniform, high-quality devices.[22, 24] The problem originates from the invasive nature of the additional electrodes that are present in multi terminal devices:[24-27] hot electrons can easily leak out through these electrodes, which effectively act as heat sinks, causing a very inhomogeneous temperature profile. Adsorbates still move around upon current annealing, but in such a temperature profile, they are just redistributed over the flake (depending on which contacts are used to current anneal the device) without being removed completely from the area of graphene probed in the transport experiments. The situation is illustrated in Figs. 1a and 1b, which show the result of transport measurements performed sequentially on a device fabricated in our laboratory (see the inset for the device configuration), after that two different pairs of contacts were used to perform current annealing. It is apparent that when measuring the resistance as a function of gate voltage ($V_G$), a sharp peak around charge neutrality is observed only when the measurement is performed with the same two contacts used to current anneal the device. This observation, which is representative of what is typically observed, directly demonstrates the ineffectiveness of current annealing when multiple metallic electrodes are invasively connected to a suspended graphene flake.

As a simple strategy to minimize the negative influence of invasive metallic electrodes, we realized suspended graphene structures in which the contacts are spaced away from the part of the device probed in the transport experiments. To this end, it is sufficient to etch broad constrictions between the metal contacts and the "active" part of the device (see the insets of Fig.

1c for a schematic drawing and Fig. 1d for a scanning electron micrograph of an actual device).[29,30] In such geometry, current annealing is achieved similarly to what is done in a two-terminal configuration by connecting contact 2 and 3 together and using them to send a current to contact 1 and 4, also connected together (i.e., multiple leads are used as a single contact to perform the annealing process; see Supporting information for more details). We have followed this strategy to realize several bilayer graphene devices and, as we now proceed to illustrate, the process results in the clean, homogeneous multi-terminal suspended devices (with a yield comparable to that of conventional two-terminal devices), enabling measurements that are not possible in a two-terminal geometry.[31]

A first indication of the high quality of our multi-terminal suspended graphene structures is obtained by measuring the resistance $R_{\alpha,\beta\text{-}\gamma,\delta}$ as a function of gate voltage $V_G$, using different contact configurations. Here, $R_{\alpha,\beta\text{-}\gamma,\delta} \equiv V_{\gamma,\delta}/I_{\alpha,\beta}$ represents the resistance obtained by measuring voltage between contact $\gamma$ and $\delta$ ($V_{\gamma,\delta}$), while sending current from $\alpha$ to $\beta$ ($I_{\alpha,\beta}$). As it is apparent in Fig. 1c, the measured resistance always exhibits a sharp peak centered around the same value of $V_G$ (= $V_{CNP}$) irrespective of the specific configuration (see Supporting Information for data measured on another device). Additionally, when plotted on a logarithmic scale, the conductance as a function of carrier density $n = \alpha \times (V_G - V_{CNP})$ (the gate capacitance $\alpha \approx 5.29 \times 10^9$ cm$^{-2}$V$^{-1}$, has been extracted from the evolution of Landau levels with increasing $B$, see below) starts increasing already for $n > 10^9$ cm$^{-2}$ (Fig. 1d; the two curves correspond to two different devices). These observations all point to an excellent device uniformity and high quality.[22]

From these measurements, we also estimate the carrier mobility $\mu = \sigma/(n \cdot e)$ and the mean-free path by extracting the conductivity $\sigma = (1/R_{1,2\text{-}4,3}) \cdot (L/W)$ from the longitudinal resistance $R_{1,2\text{-}4,3}$ (black solid line in Fig. 1c; $W = 1$ μm and $L = 1.7$ μm are the width and length of our devices).[32] We find that at a charge density of $n \approx \pm 5 \times 10^9$ cm$^{-2}$ ($|V_G - V_{CNP}| \approx 1$ V) which is well out of the low-density regime where the conductance is independent of $n$ (see Fig. 1d), the mobility $\mu$ is as high as 600,000-800,000 cm$^2$V$^{-1}$s$^{-1}$ (see quantum Hall data below for an independent estimate of $\mu$ consistent with these values). When extracting the mobility at larger density, we find even larger values. Care needs to be taken, however, because already at $n \approx \pm 5 \times 10^9$ cm$^{-2}$, the mean-free path $l_m = (h/2e^2) \cdot (\sigma/k_F) \approx 500\text{-}700$ nm extracted using Drude's formula starts to be comparable to the device dimensions.[22, 32] As we will discuss in more detail later, at carrier density values larger than $10^{10}$ cm$^{-2}$, the mean-free path exceeds the device size and the carrier motion becomes fully ballistic. In this regime, the concept of mobility, valid for diffusive transport, has no meaning on the length scale of our structures.

Before addressing the fully ballistic transport regime, we look at the quantum Hall effect to characterize the device further and gain additional confidence about its quality. Fig. 2a shows a color-coded map of the longitudinal resistance ($R_{1,2\text{-}4,3}$) measured as a function of $V_G$ and $B$ (at low magnetic field, $B < 0.5$ T; data taken at $T = 250$ mK). The multi-terminal geometry allows us to measure the magneto-resistances in both longitudinal ($R_{1,2\text{-}4,3}$) and transverse ($R_{1,3\text{-}4,2}$) configurations as shown in Fig. 2b for a fixed $V_G = -50$ V. We consistently find minima in the longitudinal resistance at a sequence of filling factor $\nu (\equiv nh/eB) = \pm 4 \times m$ (with integer $m$), accompanied by the plateaus in transverse resistance with values of $1/4m \times h/e^2$, as expected for a bilayer graphene.[33, 34] In addition, minima at $\nu = \pm 2$ (pointed to by white arrows in Fig. 2a) originating from the broken-symmetry states are also visible.[16, 35] In all cases, the value of the quantized transverse resistance plateaus (up to $\nu = 36$) matches with the value expected from the filling factor at which each plateau occurs (see Supporting information for the data at low filling factors).[26, 27] What is particularly remarkable is the low magnetic field values at which these phenomena are already visible. For instance, the broken symmetry states appear already at $B \approx$ 200 mT, whereas usually much larger values of magnetic field are needed.[35] Similarly, Fig. 2c shows the evolution of the plateau at $\sigma_{xy} = 4e^2/h$ when lowering the field ($\sigma_{xy}$, the transverse component of the conductivity tensor, is obtained from the longitudinal and the transverse resistances) which is clearly visible at $B$ as low as $B^* = 30$ mT. As well-developed quantum Hall plateaus require $\mu \cdot B^* \gg 1$, this finding implies that $\mu \gg 1/B^* \approx 300,000$ cm$^2$V$^{-1}$s$^{-1}$, consistent with the mobility extracted above from the measurement at $B = 0$ T.

We now discuss the negative multi-terminal resistance that appears at carrier densities $n > 10^{10}$ cm$^{-2}$, when measuring $R_{1,4-2,3} = V_{2,3}/I_{1,4}$ (Fig. 2d). The observation of negative resistance in this configuration implies that the carriers injected from contact 1 have higher probability to reach contact 3 than contact 2, even though contact 3 is further away from the injecting electrode. This is possible only if electrons propagate ballistically, since in the diffusive regime they would unavoidably reach the contact closer to the injection point with higher probability. The phenomenon is well established as it has been observed routinely in the past in high quality two-dimensional electron gases hosted in GaAs-based heterostructures,[36, 37] and –more recently– in ballistic samples made of graphene mono-layer encapsulated by hBN crystals.[6] In our devices, therefore, we observe a transition from diffusive to ballistic transport, i.e. positive to negative resistance, when the carrier density is increased (Fig. 2d). This is physically consistent with the result of our device characterization discussed above, indicating that the mean free path increases with increasing density, and eventually becomes larger than the device size (see the red dotted line in Fig. 2d).

Upon the application of a small perpendicular magnetic field, the negative multi-terminal resistance is suppressed, leaving a small background of a positive sign (see Fig. 3a). The effect is expected and has a classical origin.[36, 37] It is due to the Lorentz force bending the trajectories of the carriers injected from contact 1, so that they do not reach contact 3 any more: already at a small magnetic field, the deflected carriers hit the disordered edges of the sample, scattering into all directions and eventually entering contact 2 or 3 with almost equal probability (which is why the multi-terminal resistance $R_{1,4-2,3}$ nearly vanishes). In this regime, the effect of the magnetic field depends on the ratio between the cyclotron radius $r_c \equiv h/2\pi e \cdot k_F/B$ and the device dimensions. Therefore, upon increasing the gate voltage, the magnetic field needed to suppress the negative resistance is expected to scale proportionally to $k_F = |\pi n|^{1/2}$. We find that this is indeed the case: Fig. 3b shows a color plot of $R_{1,4-2,3}$ as a function of $B$ and $k_F$, in which it is clear that the sign change occurs at magnetic field values increasing linearly with $k_F$ (the magnetic field dependence of the zero crossing is summarized in Fig. 3c). Finally, note also how at very low $k_F$ (~$10^5$ cm$^{-1}$ or smaller), no negative resistance is observed. This is because at low carrier density the mean-free path is smaller than the device dimensions and the electron motion becomes diffusive. Additionally, at small values of $k_F$ the Fermi wavelength $\lambda_F = 2\pi/k_F \approx 600$ nm becomes comparable to the width of the etched constrictions connecting the central part of the device to the metal contacts, and the quasi-classical description breaks down.

As a final point, we discuss the temperature dependence of the negative resistance (Fig. 4a). It is apparent that upon increasing temperature the multi-terminal resistance $R_{1,4-2,3}$ remains negative, but only at increasingly larger values of gate voltage, i.e. the crossing point from positive to negative resistance shifts to larger values of carrier density $n$ (indicated by empty circles in fig. 4a). This behavior originates from the increase in mean-free path $l_m$ upon increasing carrier density: as scattering from phonons becomes stronger at higher temperature,[32, 38] a larger density of carriers is needed to maintain the mean-free path sufficiently long to observe ballistic transport. We find that the density at the crossing point where the multi-terminal resistance changes sign, $n \equiv n_c$, increases quadratically with temperature (see Fig. 4b). Since the crossing from positive to negative multi-terminal resistance occurs for a fixed value of the mean-free path relative to the device dimensions, these measurements can be used to extract on the temperature and the density dependence of the scattering time $\tau$ ($= l_m/v_F$, with the Fermi velocity $v_F$ given by $\hbar k_F/m \propto |n|^{1/2}$ with effective mass $m$). Specifically, if we consider a generic dependence of $\tau$ on $n$ and $T$ given by $\tau(n, T) \propto n^\alpha \cdot T^\beta$, we find that at the crossing point $const = \hbar k_F/m \cdot \tau = \hbar|\pi n_c|^{1/2}/m \cdot \tau(n_c, T) = (\hbar \pi^{1/2}/m) \cdot n_c^{\alpha+1/2} \cdot T^\beta$. By imposing the condition that $n_c$ is quadratic in $T$ as observed in Fig. 4b experimentally, we find $2\alpha + \beta + 1 = 0$, which for $\alpha = 1/2$ and $\beta = -2$ (corresponding to $\tau \propto n^{1/2} \cdot 1/T^2$) agrees with recent theory for suspended graphene bilayers[38] predicting a $1/T^2$ dependence of $\tau$ (this theory predicts various regime for the density dependence of $\tau$, depending on the experimental conditions, e.g., whether the device is under strain).[38]

In conclusion, we have demonstrated a scheme to realize multi-terminal, suspended graphene structures, which can be current annealed successfully to obtain uniform and clean devices. These devices enable new experiments not possible in a two-terminal configuration, such as the independent determination of the longitudinal and transverse magneto-resistance, and the direct observation of ballistic transport through the measurement of a negative multi-terminal resistance. In the future, these devices will prove useful to explore the properties of graphene multi-layers of different thickness, and to investigate new broken symmetry states that are being discovered in graphene-based systems.[17, 18, 21, 39-41]

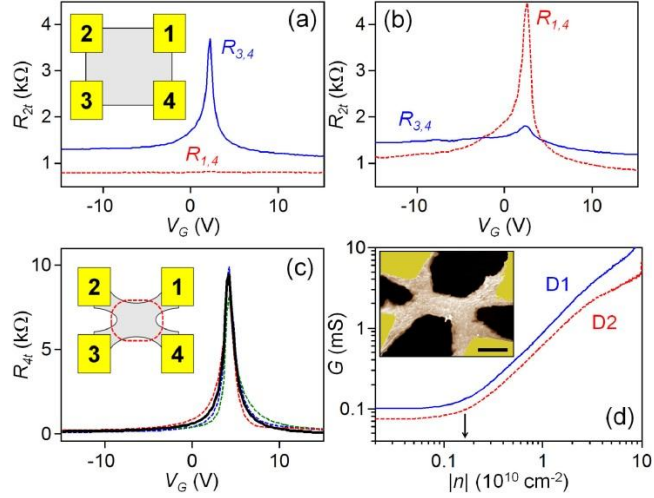

**Figure 1.** Non-invasive metal contacts are essential for successful current annealing. (a) $V_G$-dependence of $R_{3,4}$ (blue solid line) and $R_{1,4}$ (red dotted line) after that the device has been annealed by passing a large current from 3 to 4 (the inset shows the device structure with the contacts –labeled by numbers– directly connected to the graphene flake). A resistance peak is only seen in the measurements through the same contacts used to current anneal. (b) $R_{3,4}(V_G)$ and $R_{1,4}(V_G)$ measured after a second annealing step done by passing a large current between contacts 1 and 4: the resistance peak in $R_{3,4}(V_G)$ has been suppressed, while a clear peak is visible in $R_{1,4}(V_G)$. (c) Four-terminal resistances of a suspended bilayer device with the graphene etched to separate the metallic contacts from the part of the flake probed by the transport measurements (see schematics in the inset; the red broken line delimits the "active" part of the device probed in four-terminal measurements). In all measurement configurations the resistance peak is equally narrow and centered around the same gate voltage (the black solid line represents the longitudinal resistance $R_{1,2-4,3}$ used to calculate the $\mu$ and $l_m$ in the main text). (d) Longitudinal

conductance ($G = 1/R_{1,2-4,3}$) of two devices, D1 (blue solid curve) and D2 (red dotted curve), in a logarithmic scale as a function of log($n$). The arrow indicates very low level of density fluctuations, about $10^9$ cm$^{-2}$, in both devices. Inset: false-color scanning electron microscope image of a device taken before annealing (scale bar is 1 μm long). All measurements were performed at $T$ = 4.2 K.

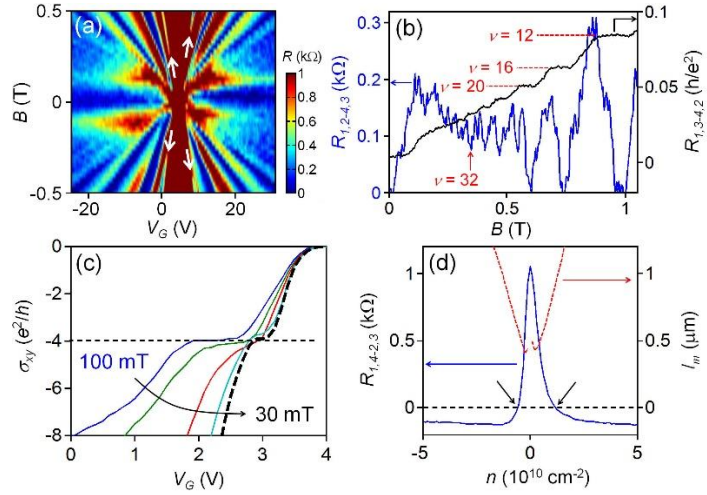

**Figure 2.** (a-c) Integer quantum Hall effect of suspended bilayer graphene measured at $T$ = 250 mK. (a) Color-coded longitudinal resistance $R_{1,2-4,3}$, plotted as a function of $V_G$ and $B$ showing clear integer quantum Hall effect of non-interacting carriers (at a sequence of filling factors $\nu = \pm 4 \times m$, with integer $m$), as well as broken symmetry states at $\nu = \pm 2$ (pointed to by white arrows). (b) Concurrent appearance of dips in the longitudinal resistance $R_{1,2-4,3}$ (blue curve) and plateaus in the transverse resistance $R_{1,3-4,2}$ (black curve; measurements taken at $V_G$ = -50 V). (c) the quantized plateau in the Hall conductivity ($\sigma_{xy}$) at $4e^2/h$ survives down to $B$ = 30 mT ($\sigma_{xy} = -R_{1,3-4,2}/[R_{1,3-4,2}^2 + R_{1,2-4,3}^2 \cdot (W/L)^2]$, with $W/L \approx 0.6$ taken from the device dimensions). (d) The multi-terminal resistance $R_{1,4-2,3} = V_{2,3}/I_{1,4}$ (left y-axis) as a function of $n$ measured at $T$ = 4.2 K, becomes negative for $n > \sim 10^{10}$ cm$^{-2}$ as indicated by black arrows. The red dotted line represents

the mean-free path $l_m$ (right y-axis) estimated from the longitudinal resistance $R_{1,2-4,3}$ (black solid line in Fig. 1c), which increases with increasing carrier density.

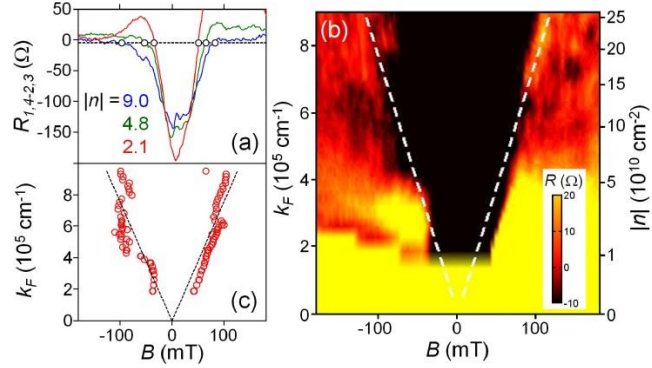

**Figure 3.** Magnetic field dependence of the negative multi-terminal resistance measured at $T = 2$ K. (a) Magneto-resistance $R_{1,4-2,3}$ measured on the hole side ($n < 0$) for different carrier densities, as indicated in the legend (in units of $10^{10}$ cm$^{-2}$). The empty circles indicate the crossing points at which the negative resistance vanishes. (b) Color-coded $R_{1,4-2,3}$ plotted as a function of $B$ and $k_F$ (on the left y-axis; corresponding values of density are shown on the right y-axis). The black region at the center of the figure represents the range of $B$ and $k_F$ in which the $R_{1,4-2,3}$ remains negative. The boundary of this region corresponds approximately to a line (white dashed lines). The linear relation between $B$ and $k_F$ at the point where the multi-terminal resistance vanishes is more precisely illustrated in panel c, where the position in $k_F$ of the crossing points (red empty circles) is plotted as a function of $B$ (see the good agreement with the linear relation between $k_F$ and $B$, shown by the black broken lines).

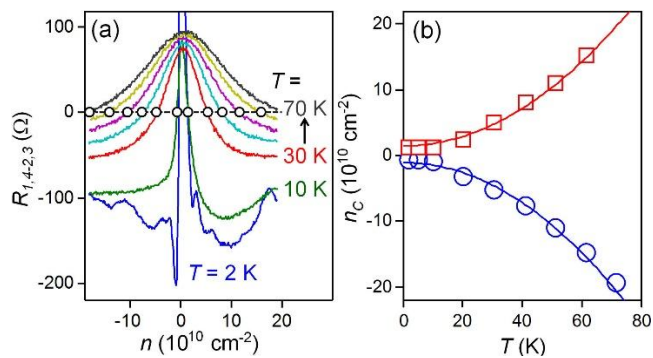

**Figure 4.** Temperature dependence of the negative multi-terminal resistance. (a) Multi-terminal resistance $R_{1,4-2,3}$ plotted as a function of carrier density for different temperatures, $T$ = 2, 10, 30, 40, 50, 60, and 70 K (from bottom to top). Empty circles indicate the crossing points at density $n = n_c$, at which the negative resistance vanishes. (b) The temperature dependence of $n_c$ (empty symbols) is well described by a quadratic function (continuous lines).

## ASSOCIATED CONTENT

**Supporting Information**. Comparable results taken from device D2, quantum Hall data at low filling factors, and more details about annealing process are provided. This material is available free of charge via the Internet at http://pubs.acs.org.

## AUTHOR INFORMATION

**Corresponding Author**

*E-mail: Alberto.Morpurgo@unige.ch.

**Notes**

The authors declare no competing financial interest.

## ACKNOWLEDGMENT

We gratefully acknowledge A. Ferreira for technical support; A. Grushina and Y. Lisunova for scanning electron microscope image of the device. Financial support from the SNF, NCCR MaNEP, and NCCR QSIT is also acknowledged.

# Supporting Information for

# "High-quality multi-terminal suspended graphene devices"

*Dong-Keun Ki and Alberto F. Morpurgo\**

Départment de Physique de la Matiére Condensée (DPMC) and Group of Applied Physics (GAP), University of Geneva, 24 quai Ernest-Ansermet, CH-1211 Geneva 4, Switzerland.

\*e-mail: Alberto.Morpurgo@unige.ch

**1. Data from a second device D2**: We show data from a second device D2 made of graphene bilayer, whose quality is comparable to that of device D1 (i.e., the device discussed in the main text). As shown in Fig. S1a, device D2 is also annealed successfully, to achieve high homogeneity in carrier density. Using the same procedure followed for device D1, also for device D2 we estimate a large value of the mobility $\mu \approx 660{,}000$ cm$^2$V$^{-1}$s$^{-1}$ and of the mean-free path $l_m \approx 530$ nm at $n \approx 5\times10^9$ cm$^{-2}$ on the hole side. The high device quality is further confirmed by the integer quantum Hall effect (Fig. S1b) appearing at low magnetic fields (white arrows indicate broken symmetry states which start to occur at low field, about 0.5 T). Fig. S1c shows that due to the high quality (and having a geometry nearly identical to that of device 1), device D2 also exhibits a negative resistance $R_{1,4\text{-}2,3}$ already at fairly small values of carrier density.

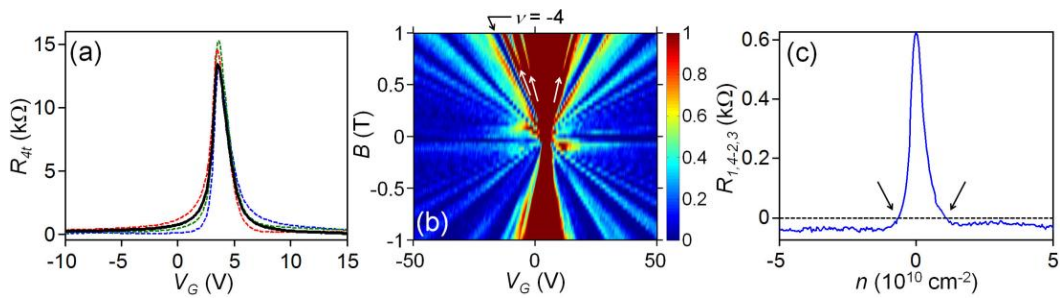

**Figure S1**. (a) Four-terminal resistance $R_{4t}$ as a function of $V_G$ measured at $T = 4.2$ K in different configurations. Here, the solid line represents the longitudinal resistance $R_{1,2\text{-}4,3}$ whose conductance is plotted in logarithmic scale in Fig. 1d of the main text. (b) Fan-diagram $R_{1,2-4,3}(V_G, B)$ measured at $T = 0.25$ K. The black arrow indicates the minimum in resistance at $\nu = -4$, which survives at magnetic field values well below $B = 0.1$ T. The white arrows point to symmetry broken states. Panel (c) shows the negative multi-terminal resistance $R_{1,4-2,3}$ at $T = 4.2$ K that is observed in device D2, analogously to the case of device D1 discussed in the main text (the black arrows indicate the crossing from positive to negative resistance as the carrier density is increased).

**2. Clear quantum-Hall quantization at low filling factors**: Our devices exhibit almost perfect quantization of transverse Hall resistances as shown in Fig. 2b of the main text. In Fig. S2, we further shows that it is also true for the states at low filling factors down to $\nu = 1$.

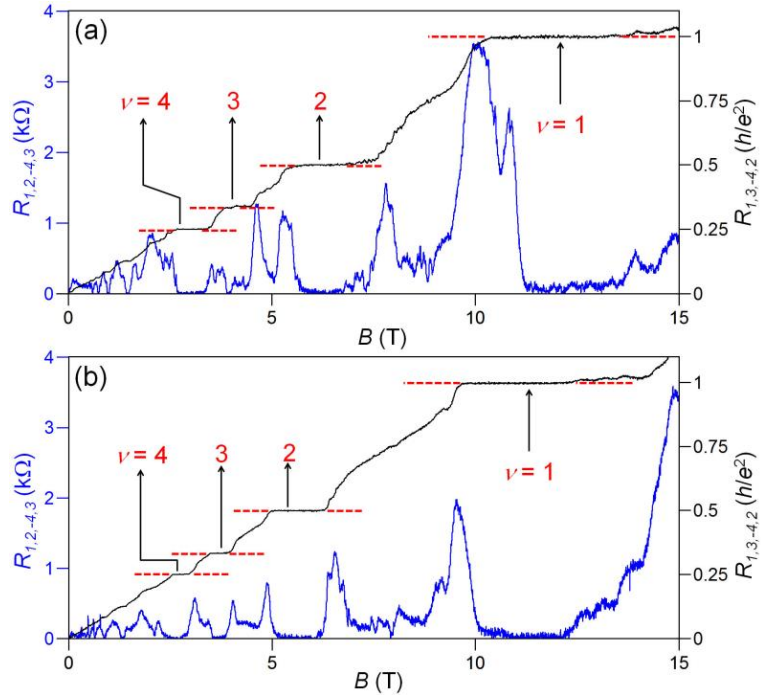

**Figure S2**. Magneto-resistance $R_{1,2-4,3}$ (blue curves; left y-axis) and $R_{1,3-4,2}$ (black curves; right y-axis) taken at $V_G = -50$ V ($T = 0.25$ K) from both devices D1 (a) and D2 (b). Exact quantum Hall quantization at $\nu = 1, 2, 3,$ and 4 is clearly seen.

**3. Current annealing process**: The annealing process is essential to achieve high quality suspended graphene devices. In our multi-terminal devices, by short-circuiting contacts in pairs, we can use the same scheme employed to anneal two-terminal devices, whose details are provided below.

A first important point is to perform the current annealing in a voltage-biased configuration (i.e., by fixing the voltage $V_b$ across the device) rather than in a current-biased configuration (see the scheme drawn in Fig. S3a). A voltage-biased scheme provides better control, as it prevents the power dissipated through the device, $P = I_b^2 R_s = V^2/R_s$, to increase when the resistance ($R_s$) increases (as it normally does during successful annealing). Therefore, by reducing excessive power dissipation, a voltage biased scheme eliminates one of the causes of device failure. In addition, it is important to monitor the current-voltage characteristics of the device during the process (see Fig. S3b, for instance), as the resulting information is used to decide when to stop increasing $V_b$ and remain at constant applied bias, and when to decrease $V_b$ to zero. After decreasing $V_b$ to zero, the device quality is investigated by measuring the gate-voltage dependence of the resistance (Fig. S3c). It normally happens, as shown in Figs. S3b-c, that the current annealing process has to be repeated several times before the quality of the device is satisfactory. The precise values of maximum current and voltage during the annealing depend on the device dimensions, but they are typically of the order of 1 milli-ampere per micron of width of the graphene layer, corresponding to a bias voltage of approximately 2-3 volts. As for the yield of successfully annealed devices, we found that this critically depends on the specific batch of samples. In some batches, the yield was superior to 50 %, but in others, it was poorer with most devices failing upon annealing. The precise mechanism for these batch-to-batch fluctuations is not currently clear.

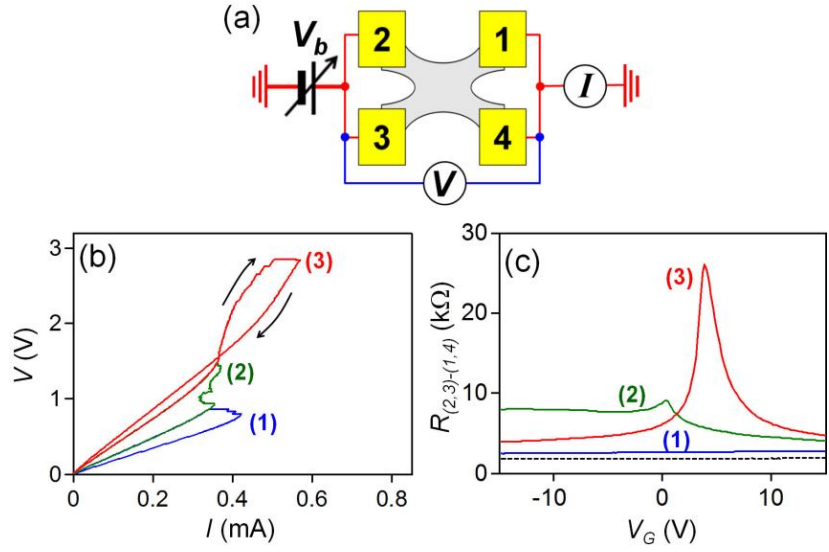

**Figure S3**. (a) Scheme for the annealing process. The voltage $V_b$ is applied across the device using a Keithley 2400 source-meter (which measures the current $I$ flowing through the device at the same time), and the actual voltage drop across the device $V$ is measured using a HP digital multi-meter (during current annealing, the device is mounted in the cryostat at 4 K, and all the lines are low-pass filtered with a cut-off frequency of ~ 1 kHz). (b) Current-voltage characteristics of the device recorded during three subsequent annealing processes, with the data taken in the order marked by the corresponding numbers. (c) Gate-voltage dependence of the resistance measured after each annealing attempt are shown. The black dotted line, which shows nearly no gate voltage dependence, corresponds to the data taken before annealing.